\documentclass[epj]{svjour}
\usepackage{graphics}
\usepackage{url}
\begin{document}
\title{Diagnostics of plasma in the ionospheric D-region: detection and study of different ionospheric disturbance types}
\author{Aleksandra Nina\inst{1}\thanks{\emph{Present address:} A. Nina, Institute of Physics, University of Belgrade, Pregrevica 118,11080 Belgrade, Serbia, email:  sandrast@ipb.ac.rs}, Vladimir  M. \v{C}ade\v{z}\inst{2}, Luka \v{C}. Popovi\'{c}\inst{2} \and Vladimir A. Sre\'{c}kovi\'{c}\inst{1}
}                     
\offprints{}          
\institute{Institute of Physics, University of Belgrade, Pregrevica 118, 11080 Belgrade, Serbia \and Astronomical Observatory, Volgina 7, 11060 Belgrade, Serbia}
\date{Received: date / Revised version: date}
%
\abstract{Here we discuss our recent investigations of the ionospheric plasma by using very
low and low frequency (VLF/LF) radio waves.
We give a review of how to detect different low ionospheric reactions (sudden ionospheric disturbances) to various terrestrial and extra-terrestrial events, show their classification according to intensity and time duration, and present some methods for their detections in time and frequency domains. Investigations of detection in time domain are carried out for intensive long-lasting perturbations induced by solar X-ray flares and for short-lasting perturbations caused by gamma ray bursts. We also analyze time variations of signals used in the low ionospheric monitoring after earthquake events. In addition, we describe a procedure for the detection of acoustic
and gravity waves from the VLF/LF signal analysis in frequency domain. The research of the
low ionospheric plasma is based
on data collected by the VLF/LF receivers located in Belgrade, Serbia.
\PACS{
      {52.25.Jm}{Ionization of plasmas}   \and
      {94.20.Bb}{Wave propagation} \and
      {94.20.de}{D region} \and
      {92.60.hh}{Acoustic gravity waves, tides, and compressional waves}
     } 
} 
\authorrunning{Nina et al.}
\titlerunning{Diagnostics of plasma in ionospheric D-region}
\maketitle
\section{Introduction}
\label{intro}

The ionosphere is the part of the atmosphere located between about 50 km and 1000 km where the charged particles significantly influence its physical and chemical properties \cite{mit74,bot07}. For this reason the study of plasma properties and all processes with a focus on charged particle production plays a crucial role in its understanding. In addition to scientific importance \cite{rat12,jil13,hai16}, studies of the ionospheric plasma  and  the dynamics of perturbations induced therein  can, for example,  be of great practical  significance in fields related to telecommunications \cite{ayu15} and may also contribute to a better insight into features related to elementary disasters like earthquakes \cite{hay07}.

The ionosphere being a  part of the terrestrial outer layer is constantly exposed to many influences coming from the outer space in addition to those occurring  in the Earth's layers which all affect its dynamics \cite{jak05,cli09,sil16}. Consequently, the physical properties of the atmosphere (density, temperature etc.) are time and space dependent \cite{nin14,baj15} which justifies the use of monitoring variations of atmospheric parameters for indirect detection of different phenomena both of cosmic and terrestrial origin \cite{mau14,zha15,kum16}.

Generally, the phenomena that affect the local environment cause different reactions of its constituents. These responses vary in intensity, duration and location of the perturbation, which requires application of various observational setups and techniques for their detection. Some of the most important measurement characteristics are the distance between the experimental setup and observed area,  observed altitude range, time resolution and sensitivity. Based on this, one can divide the observation methods in various ways:
\begin{itemize}
  \item Based on distance between the experimental setup and observed area. There are two types of atmospheric monitoring: in situ (by rockets and satellites/space $\,$ probes) and remote sensing (by satellites, radars, iono\-sondes, very low and low frequency (VLF/LF) emitters and receivers). In the first case, the instruments are placed at the location that is being observed while the remote sensing observations are based on detection of signals emitted by transmitters located at some distance away from the receivers and the area which is being monitored. Data obtained by remote sensing techniques are less precise than those by in situ measurements but they cover a significantly broader observational area.

  \item Based on the observed altitude range. Thus, LIDAR (Light Detection and Ranging) \cite{bin15} is used for observations of the atmosphere at altitudes of a few kilometers, balloons are enforceable for measurements at about 30 km \cite{sch15} while radar, rocket, and measurements  by radio signals \cite{str10,tod16,cha14,ver16} can be used for observations of the ionosphere.
  \item Based on time resolution, taking advantage from the fact that the duration times of phenomena can be very different. For example, some disturbances, such as those caused by lightnings, gamma ray bursts (GRBs)  \cite{nin15a} and meteor passings through the atmosphere \cite{zgr02} can last up to several ms while phenomena like solar X-ray flares, coronal mass ejections, and hurricane events can perturb part of the atmosphere for periods lasting  tens of minutes, hours, or days. For this reason the time resolution of the received data must be adapted to the duration time of the observed phenomenon.
  \item Based on sensitivity, which is particularly important in the case of weak perturbations. In addition to instrument characteristics it depends on the area where signals propagate (length of the signal propagation path and plasma medium properties).
\end{itemize}

In this paper we focus our research on the lower ionosphere located between 50 km and 90 km where the dominant source of ionization under unperturbed conditions comes from the solar Ly$\alpha$ radiation (above about 70 km at daytime) which induces formation of the D-region, and cosmic rays (at nighttime and below about 70 km at daytime). 
Variations in intensity of these radiations as well as the increase of incoming X and gamma radiation fluxes in the atmosphere, induction of different types of waves, and changes in the atmospheric electric conductivity cause sudden changes in the ionospheric plasma properties. The sources of these sudden ionospheric disturbances (SIDs) have extraterrestrial  and terrestrial origins. The most important SIDs result from solar X-ray flares and lightnings while influence of other noticed phenomena like GRBs rarely induce very significant SIDs. These perturbations have various properties reflected in their duration and intensity which can further be used for their classification.

The aim of the paper is to present methods of SID detection and their differences depending on causing events (terrestrial and extra-terrestrial) based on time and frequency domain analysis. We point out the importance of certain characteristics of the low ionosphere monitoring in study of its non-periodic, local, short-term, and weak reactions. Keeping in mind that SIDs can be classified according to their duration time as short-term (like in the case of lightning occurrence) and long-term (like in the case of some solar influences), and based on intensity to strong (e.g.induced by solar X-ray flares) and weak (e.g. due to GRBs), we classify the relevant procedures for extraction of SIDs. This kind of investigation is important for further analyses of characteristics of plasma  parameters such as electron density, electron gain and loss rates, recombination coefficients and temperature under perturbed conditions \cite{baj15,nin12b,nin12a,sch13}. In these researches it is necessary to implement a numerical program package like the Long-Wave Propagation Capability (LWPC) \cite{fer98} and use analytical procedures like those given in \cite{nin14,sch13}.

Here we analyze the indirect detection of non-periodic phenomena and solar terminator (ST) by monitoring VLF/ LF radio waves whose propagation  depends on the low ionosphere properties and, consequently, varies with induced disturbances in this atmospheric part. The presented examples of given procedures are obtained in analyses of data collected by the AWESOME (Atmospheric Weather Electromagnetic System for Observation Modeling and
Education) \cite{coh10} and AbsPAL (Absolute Phase and Amplitude Logger) VLF/LF receivers in Belgrade, Serbia for signals emitted by the DHO, ICV, and NAA transmitters located in Germany, Italy and the USA, respectively.


Here we limit our study to ionospheric variations after and during occurrences of considered events. We point out that there are some studies which analyze prediction of particular events using ionospheric perturbation detections, but this task requires a more detailed analysis which is out of the scope of this paper but will be in focus of our upcoming investigations.

The paper is organized as follows: In Sec. \ref{sec:1} we describe our observations and experimental setup, in Sec. \ref{tda} we present time domain analysis, and in Sec. \ref{sec:frequ} we give a model for detection of SIDs using the frequency analysis. Results of our research concerning detections of SIDs using VLF signal analyses in time and frequency domain are presented in Sec. \ref{sec:res}, and, finally, a short summary of this study is given in Sec. \ref{sec:sum}.

\section{Observations and experimental setup}
\label{sec:1}
As said in Introduction, this study is focused on ground based measurements of the VLF/LF radio waves (3 kHz - 30 kHz and 30 kHz - 300 kHz frequency domains) which are reflected in the Earth-ionosphere waveguide.
This met- hod is based on the fact that the considered signals propagate through the low ionosphere which  affects characteristics of their propagation and, consequently, the shapes of registered VLF/LF wave variations in real time, indicating the presence of non-stationary physical and chemical conditions in the perturbed medium along the VLF/LF wave trajectories (for details see, for example \cite{nin11,nin12b}).
Namely, perturbations make the local electron density and, consequently, the height of the wave reflection, time dependent \cite{nin12a} which further alters the VLF/LF wave trajectory and causes the registered wave amplitude and phase to be time dependent.

\begin{figure}
\resizebox{0.45\textwidth}{!}{%
  \includegraphics{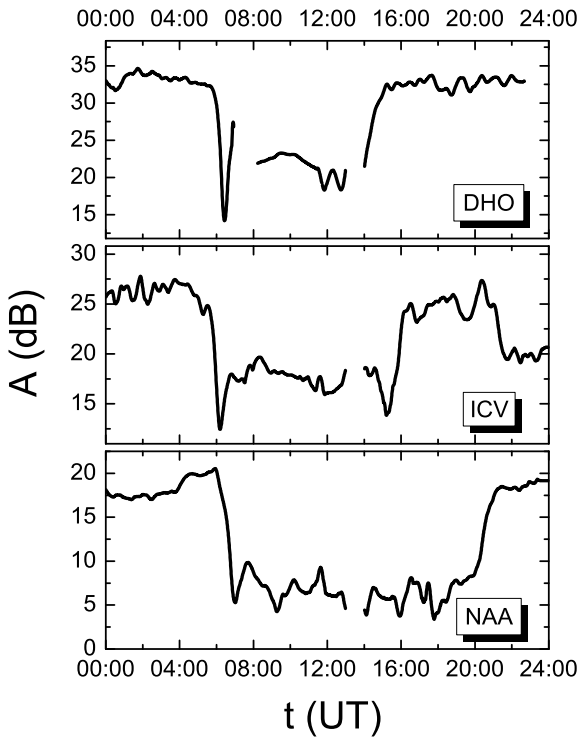}
}
\caption{Time-evolutions of the amplitude of signals emitted by the DHO (upper panel), ICV (middle panel) and NAA (bottom panel) transmitters and recorded by the Belgrade AWESOME VLF/LF receiver on December 21, 2010.}
\label{fig:dan3signala}
\end{figure}

\begin{figure*}
	\begin{center}
\resizebox{0.85\textwidth}{!}{%
  \includegraphics{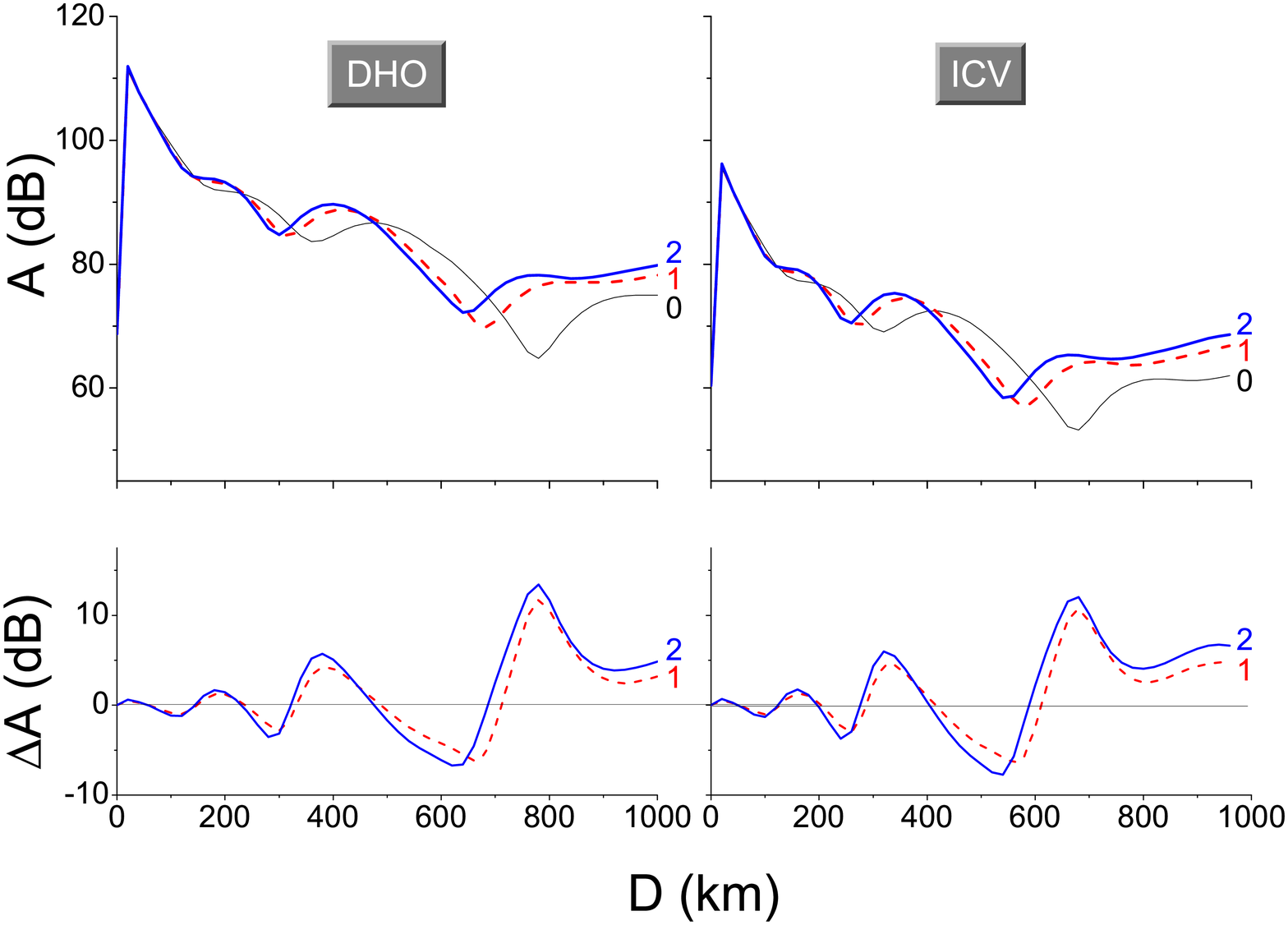}
}
\caption{Simulated amplitudes (upper panels) and  their changes (bottom panels) relative to the initial quiet ionospheric state of the VLF signals versus distance from the DHO (left panels) and ICV (right panels) transmitters that emit them during the quiet condition "0" and in two perturbed stages "1" and "2" induced by the solar X-ray flare occurred on May 5, 2010.}
\end{center}
\label{ModMin}
\end{figure*}

The global experimental setup for the VLF/LF monitoring technique consists of numerous
transmitters and receivers distributed worldwide  that enable observations of a large part of the low ionosphere and detections of local plasma perturbation patterns in the D-region. A number of receivers are incorporated in some international networks like AWESOME \cite{coh10},  SAVNET \cite{rau09} and AARDDVARK \cite{cli09}. A very important characteristic of this technique for ionospheric monitoring is a continuous emission and reception of radio signals with a very good time resolution (it can be 10 $\mu$s) allowing  detections of sudden, and, consequently, non precisely predicted, events, as well as detection of short-term ionospheric reactions.

In this work we present data recorded by the VLF/LF AWESOME and AbsPAL receivers located in the Institute of Physics in Belgrade, Serbia. These receivers operate since 2004 and 2010, respectively. Here we study connections of SIDs with three specific events for time-domain: GRBs (2009-2012), solar X-flare in 2010 and earthquake in Serbia in 2010, as well as connections with ST in 2010 for the frequency domain analysis.

We consider the 23.4 kHz, 20.27 kHz and 24 kHz signals emitted by the DHO (Germany), ICV (Italy) and NAA (USA) transmitters, respectively, with time resolutions of 0.02 s (for the analysis of short-term SIDs, such as those induced by GRBs), 1 s (for the analysis of perturbations induced by solar X-ray flares) and 1 min (for the analysis of long-term SIDs, possibly induced by earthquakes). The emission power of these signals is 800 kW, 20 kW and 1000 kW, respectively, and they are transmitted as can be seen in Fig. \ref{fig:dan3signala} showing the amplitudes recorded by the Belgrade AWESOME receiver on December 21, 2010. The detection breaks occurring in time intervals 7-8 UT and 13-14 UT
are of a pure technical nature related to the DHO transmitter
being off-air, and the procedure of how the collected data are
preprocessed and archived \cite{coh10}, respectively.

\section{Time domain analysis}
\label{tda}

Characteristics of a detected signal by a particular device for the VLF/LF radio wave acquisition are time dependent and connections of their signatures with ionospheric perturbers are very hard to track. There are two main reasons for that:
\begin{itemize}
  \item First, as it was anticipated in Sect.\ref{intro}, the plasma located in the low ionosphere is simultaneously exposed to influences of numerous natural and artificial events. Consequently, the recorded signal characteristics which indirectly reflect ionospheric plasma properties are subject to noise and different tendencies which become of prime importance in detection of particularly weak perturbations.
  \item Second, in addition to periodical and sudden variations of ionospheric plasma conditions, characteristics of signals like mutual locations of the transmitter and receiver, power of transmitted signal, and the geographical area through which the signal propagates affect the recorded signal properties. Namely, the intensity of the received signal amplitude depends on the emission power and on the distance between the transmitter and receiver. 
  In the case of emitter power, a more intense emission induces a larger amplitude of the received signal than emitted signal with lower power. On the other hand, the influence of the transmitter-receiver distance on the considered relationship is not so simple. This can be visualized using simulations of signal propagation within the Earth-ionosphere waveguide by the LWPC numerical model developed by the Naval Ocean Systems Center (NOSC), San Diego, USA \cite{fer98}. In Fig. \ref{ModMin}, upper panels, we show simulated  amplitude of signals emitted by the DHO transmitter (Germany) with the emission power of 800 kW and by the ICV transmitter (Italy) with the emission power of 20 kW at the ground in directions toward the Belgrade VLF/ LF receiver. We analyze three moments in period before and during the solar X-ray flare occurred on May 5, 2010.
      We take the quiet period before the flare, denoted as period "0" in the panels, as the reference dependency, while the periods "1" and "2" correspond to perturbed stages where we estimated the ionospheric conditions, before the perturbation maximum and at the peak of the perturbation, respectively.

In this figure we can see that the amplitude intensity is weaker for the ICV signal than for the stronger DHO signal at all distances from the transmitters. Also, the dependencies between the ionospheric changes of electron density induced by the X-ray radiation increase and the VLF/LF signal amplitudes are not monotonous, e.g. a growth of the electron density ($N_e(0)<N_e(1)<N_e(2)$) does not necessarily  imply an increase of the recorded signal amplitudes.
This is visible in the bottom panels showing that the amplitude changes can be both positive and negative when the electron density is larger (cases "1" and "2") than in the case of quiet conditions "0". One can see that these changes in amplitude depend on the distance from the transmitter.


As an experimentally recorded example of different reaction on electron density increase/decrease we present amplitude variations of signals emitted by the NAA, ICV and DHO transmitters located in the USA, Italy and Germany, respectively, which were induced by the electron density increase due to the solar X-ray flare occurred on March 25, 2011 (see Fig. \ref{fig:signali3}). It is noticeable that more intensive ionization processes result in the increase of the NAA signal amplitude (upper panel), decrease of the ICV signal amplitude (middle panel) and the combined tendency of the DHO signal amplitude (bottom panel).

\begin{figure}
\resizebox{0.45\textwidth}{!}{%
  \includegraphics{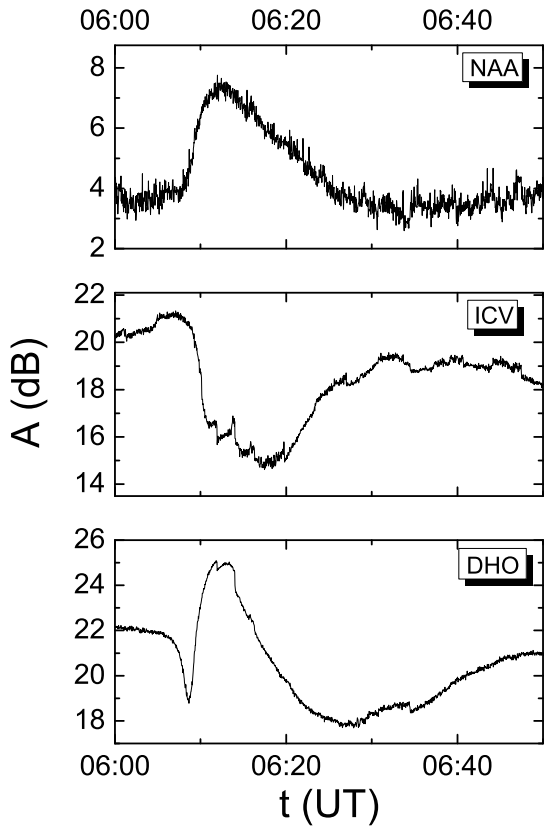}
}
\caption{Differences in the amplitude time evolutions of the signal emitted in the USA (upper panel), Italy (middle panel), and Germany (bottom panel) and received by the AWESOME VLF/LF receiver in Serbia during the influence of the solar X-ray flare occurred on March 25, 2011.}
\label{fig:signali3}
\end{figure}

\end{itemize}

\begin{figure}
\resizebox{0.45\textwidth}{!}{%
  \includegraphics{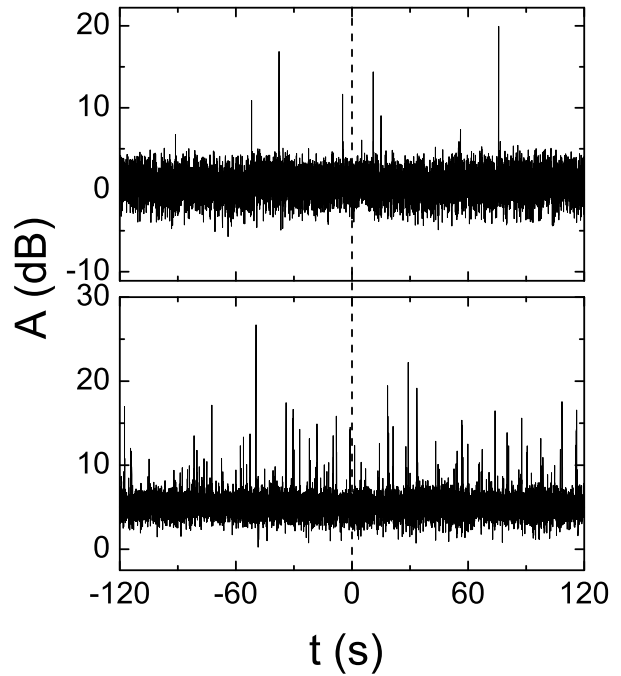}
}
\caption{Examples of the signal amplitude time evolutions during quiet (upper panel) and perturbed (bottom panel) conditions.}
\label{fig:nesigurnosti}
\end{figure}

Despite of numerous impacts on the ionospheric plasma, individual events can cause dominant influence on local plasma properties. In that case the relationship between a particular event and the corresponding SIDs (here classified as strong SIDs) can be analyzed. However, in some cases we can not extract the low ionospheric reaction induced by some considered events. There are three main reasons for that:

 \begin{itemize}
   \item The intensity of the SID is weak and it cannot be extracted from the noise in one particular case;
   \item The shape of signal variation caused by the considered event is the same or very similar to those induced by some other phenomena occurring in the same time period;
   \item The reaction does not induce clearly visible changes in recorded signal properties
 \end{itemize}
 Here, we take these reactions as weak SIDs. The study of the relationship between the considered events and relevant local plasma reactions is primarily based on statistical analyses. For these procedures it is very important that no other processes, inducing  reactions similar to those expected in the considered case, are present. For example, the statistical analysis of the short term SIDs induced by GRBs is not relevant in the case when numerous amplitude peaks exist before the satellite detection of the GRB (Fig. \ref{fig:nesigurnosti}, bottom panel).

There are different procedures for detection of weak SIDs. Here we explain the following methods:

\begin{itemize}
  \item \textbf{Extraction of amplitude peaks.}
  This technique is used and described in \cite{nin15a} where the  short duration low ionospheric reaction on a GRB is confirmed. This is based on determination of times when the peaks of the signal amplitude $A(t)$ deviate from the base curve $A_{\rm {base}}(t)$ by more than $r$ times the amplitude of noise $A_{\rm {noise}}(t)$:
   \begin{equation}
    \label{eq:1prvo}
    {
    \frac{A(t)-A_{\rm {base}}(t)}{A_{\rm {noise}}}\geq r,
    }
  \end{equation}
   and their occurrence in time bins before or after the registration of the considered events.
  \item \textbf{Comparison with relevant quiet period.} Information about the existence of SIDs can be obtained using a comparison of signal characteristics from time periods with practically the same conditions, but in absence of SIDs. This procedure is very useful in the case when there are no sudden strong variations in the signal characteristics time evolutions.
  \item \textbf{Superposed epoch technique.} This technique is applicable when  we have a weak perturbation which is not clearly visible in one particular case because of the low signal intensity but which is repeated under the same influence. For example, this method is used to detect the transmitter-induced precipitation
of the inner radiation belt electrons in the low ionosphere which is practically very small \cite{ina07b}. It is based on the averaged sum of time series of the signal amplitude. Here it is important to say that perturbations are of the same duration. Namely, in the case when the amplitude variations are short-term with respect to the considered time interval, and are not frequent, this method can not confirm them although their occurrences are in the same time period after the influence of perturber. This conclusion is obtained for a short term response of the low ionosphere on GRBs \cite{nin15a} whose detection is confirmed using the procedure for extraction of amplitude peaks.
\end{itemize}

For both types of SIDs, strong and weak, analyses of the ionospheric plasma require independent data related to detection of processes which perturb it like data collected by satellite-borne detectors for radiation coming from the outer space or ground-based optical meteor detectors.

In addition to the SID detection using the analysis of signal characteristics in time domain, SID can be discovered as intensification of waves at some frequency in the considered medium. In diagnostic of the low ionospheric plasma by radio waves, this method can be applied to electron density variations in periods when we can assume a monotonous relationship between the recorded signal characteristic and electron density.

\section{Frequency domain analysis}
\label{sec:frequ}

In addition to the SID detection based on the signals analysis in the time domain, the recorded data can be processed by appropriate techniques and further analyzed in the frequency domain in order to extract waves existing in the considered medium. These waves can be excited by natural \cite{art05,nin13a} and artificial \cite{mab16} events and, depending on their frequency and medium properties, they can be divided into several types like acoustic, gravity and planetary waves.

Here, our attention is focused on explanation of the theoretical procedure for determination of possible acoustic and gravity waves (AGWs) in the lower ionosphere and on presentation of processing for determination of the excited wave frequency from the VLF/LF signal amplitude analysis.

\subsection{Theory}
\label{AGWtheory}

Keeping in mind that typical atmosphere models give $n_n\sim 10^{21}\rm{m}^{-3}$ for the neutral particle density and only $n_p\sim 10^{8}\rm{m}^{-3}$
for charged plasma particles at heights $H_r < 90$ km where VLF/LF
radio waves are being deflected, we can assume that the electric and magnetic
effects play a negligible role in local dynamics. Consequently, standard hydrodynamic rather than
magneto-hydrodynamic (MHD) equations can be applied in analysis of the low ionospheric waves.
For this reason in our study we start therefore from the general set of hydrodynamic equations for adiabatic processes in ideal
neutral gas:
\begin{equation}
\label{e02}
\frac{\partial\rho}{\partial t}+\nabla\cdot(\rho\vec v)=0,\quad p=\rho R_aT,
\end{equation}
\begin{equation}
\label{e02a}
 \rho\frac{\partial\vec v}{\partial t}+\rho(\vec v\cdot\nabla)\vec v=-\nabla p+\rho\vec g,
\end{equation}
\begin{equation}
\label{e02b}
  \frac{\partial p}{\partial t}+(\vec v\cdot\nabla)p
=\gamma\frac{p}{\rho}\left[\frac{\partial\rho}{\partial t}+(\vec v\cdot\nabla)\rho\right].
\end{equation}
Here $\gamma=c_p/c_v=(i+2)/i$ is the ratio of specific heats for gas particle
with $i$ degrees of freedom ($i$=3,5 for a mono atomic and two atom molecules, respectively),
 $R_a=k/m_a=R/M_a$ is the
individual gas constant for molecules with particle mass $m_a$ or molar mass $M_a$, $k=1.3807\times 10^{-23}\mbox{ [J/K]}$ is Boltzmann's constant, and
$R=$8.3145 $\,$ [J/K/mol] is the universal gas constant. Other quantities in Eq.(\ref{e02})
have their usual meanings.

In what follows, we consider waves whose spatial dimensions are sufficiently small in comparison
with both the radius of  the Earth $R_E=6371$ km and any temperature inhomogeneity length
${\cal L}_T$ that can be defined on the basis of existing temperature profiles. Consequently,
the plane parallel geometry can be applied with gravitational acceleration $\vec g=-g\hat e_z$
(with $g$=9.81 m/s$^2$)
in a locally isothermal medium. Under these assumptions the atmosphere is taken to be vertically
stratified, initially in hydrostatic equilibrium, and then perturbed by harmonic waves of
small amplitude. This means that Eqs (\ref{e02})-(\ref{e02b}) can be linearized by taking each variable
$\Psi(x,y,z,t)$ as a sum of its basic state unperturbed value $\Psi_0(z)$ and a small
first order perturbation $\Psi_1(x,y,z,t)$ arising from waves, i.e.:
\begin{equation}
\label{e03}\Psi(x,y,z,t)=\Psi_0(z)+\Psi_1(x,y,z,t),
\end{equation}
where:
\begin{equation}
\label{e03a}|\Psi_1(x,y,z,t)|\ll|\Psi_0(x,y,z,t)|
\end{equation}
and
\begin{equation}
\label{e03b}\Psi_1(x,y,z,t)=\hat{\Psi}_1(z)e^{-i\omega t+i(k_xx+k_yy)}.
\end{equation}

Expressed in terms of the wavelength $\vec\lambda\equiv 2\pi(1/k_x,$ $1/k_y,1/k_z)$, our modal
analysis is restricted to atmospheric waves obeying the
conditions:
\begin{equation}
\label{e04}\lambda_x,\lambda_y\ll R_E,\quad\lambda_z\ll{\cal L}_T
\end{equation}
which is equivalent to studying plane waves in a horizontally stratified isothermal atmosphere.

Eqs. (\ref{e02}) - (\ref{e02b}), linearized with perturbations given by Eqs. (\ref{e03}) - (\ref{e03b}) and Eq.(\ref{e04}), reduce to
two equations: one for the basic unperturbed state and one for small perturbations.

The basic unperturbed state is thus described by:
\begin{equation}
\label{e04a}\frac{dp_0}{dz}+\rho_0g=0,
\end{equation}
\begin{equation}
\label{e04b}p_0=\rho_0 R_aT_0,\mbox{  with  } T_0=\rm{const},
\end{equation}
whose solution is:
\begin{equation}
\label{e06} \rho_0(z)=\rho_0(0)e^{-z/H} \mbox{    or:    }
p_0(z)=p_0(0)e^{-z/H}
\end{equation}
with $H$ being the characteristic scale-height of the isothermal atmosphere given by:
\begin{equation}
\label{e07}  H\equiv \frac{p_0(0)}{\rho_0(0)g}=\frac{v_s^2}{\gamma g},
\end{equation}
while the small perturbations are governed by:
\begin{equation}
\label{e07b}
v_s^2\omega^2\frac{d^2\hat v_{1z}}{dz^2}-\gamma g\omega^2\frac{d\hat v_{1z}}{dz}
+\left(\omega^4-k_0^2v_s^2\omega^2+k_0^2v_s^2N^2_{BV}\right)\hat v_{1z}=0.
\end{equation}
Here $\omega$ is angular frequency,  $k_0^2\equiv k_x^2+k_y^2$ is the total horizontal wavenumber,
$v_s$ is the adiabatic speed of sound defined as:
\[
  v_s^2\equiv\gamma\frac{p_0}{\rho_0}=\gamma R_aT_0=\rm{const},
\]
and  $N{_{BV} }$ is the
Brunt-V\"ais\"al\"aa frequency given as:
\begin{equation}
\label{e07a}  N_{BV}^2=(\gamma-1)\frac{g^2}{v_s^2}.
\end{equation}

Taking Eq.(\ref{e06}) into account, Eq.(\ref{e07b}) has solutions of the following form:
\begin{equation}
\label{e08}  \hat v_{1z}=C\times e^{(1/2H\pm ik_zz)}
\end{equation}
which finally yields the dispersion relation:

\begin{equation}
\label{e09}\omega^4-\left(k_0^2+k_z^2+\frac{1}{4H^2}\right)v_s^2\omega^2+k_0^2v_s^2N^2_{BV}=0
\end{equation}
that has to be satisfied for solutions  like Eq.(\ref{e08}) to exist with a non-zero integration
constant $C$.

The dispersion relation Eq.(\ref{e09}) is quadratic in $\omega^2$ which indicates the existence of two wave-modes
in the considered medium: AGW modes, also known as p- and g-modes
in stellar seismology.

Due to the boundary conditions of Eq.(\ref{e04}), it is convenient to express
the dispersion relation Eq.(\ref{e09}) in terms of wavelengths and wave frequencies
in the following way:
\begin{equation}
\label{e10}
\lambda_0^2(f)={\cal D}_0(f)\left[1+\frac{{\cal D}_2(f)}{\lambda_z^2-{\cal D}_2(f)}\right]
\end{equation}
with:
\begin{equation}
\label{e11}{\cal D}_0(f)=
\frac{v_s^2(f^2-{f_{BV}}^2)}{f^2(f^2-{f_0}^2)}, \quad
{\cal D}_2(f)=\frac{v_s^2}{f^2-{f_0}^2},
\end{equation}
and:
\begin{equation}
f_{_0}=\frac{\gamma g    }{4\pi v_s},\quad f_{_{BV}}
=\frac{N_{BV}}{2\pi},\quad \lambda_{0,z}=\frac{2\pi}{k_{0,z}},
\end{equation}
where $f_0$ and $f_{_{BV}}$ correspond to the acoustic cut-off and Brunt-V\"ais\"al\"aa frequencies, respectively.

\begin{figure}
\resizebox{0.5\textwidth}{!}{%
  \includegraphics{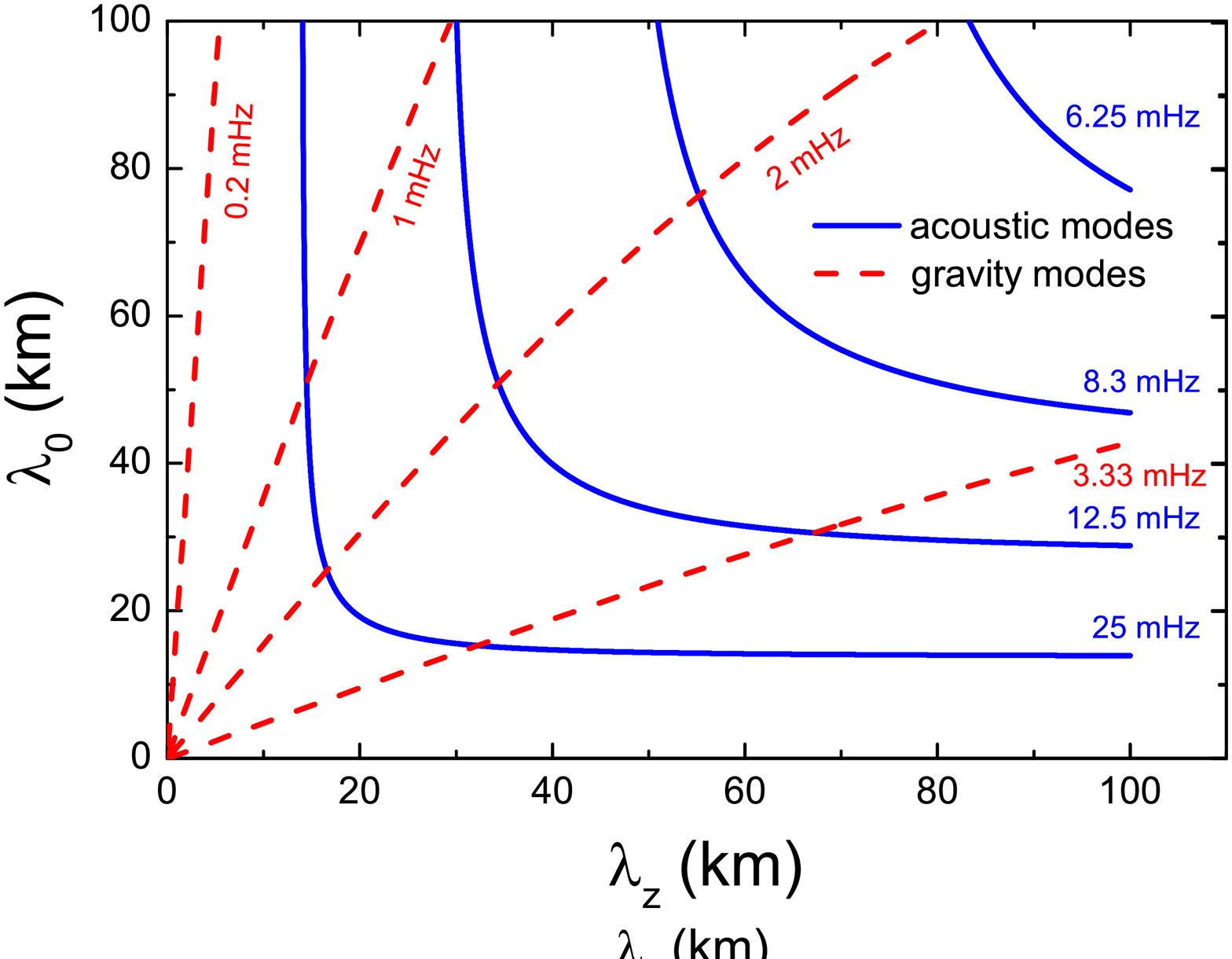}
}
\caption{Acoustic (solid lines) and gravity (dashed lines) modes for several wave frequencies.}
\label{fig:RADobeMODE}
\end{figure}

A family of hyperbolae obtained from Eq.(\ref{e10}) in a ($\lambda_0$,$\lambda_z$)-plot with $f$
being a parameter is shown in Fig. \ref{fig:RADobeMODE}. In calculations we take $T_0=200$ K as typical
temperature of the considered medium.
The dispersion relation Eq.(\ref{e10}) has two separate domains describing:
\begin{itemize}
  \item acoustic modes if $\infty>f \geq f_{_0}=0.00606$ Hz, and
  \item gravity modes for $f\le f_{_{BV}}=$ 0.00594 Hz.
\end{itemize}
There are no propagating waves with $f_{_{BV}}\le f\le f_{_0}$.

\subsection{Signal processing}

One of the procedures for frequency determination of the excited waves is based on the Fast Fourier Transform (FFT) applied on collected data for VLF/LF amplitude using equation:
\begin{equation}
\label{e01}  A_{\rm {F}}(\omega)=\frac{1}{\sqrt{2\pi}}\int_{-\infty}^{+\infty}{\rm{e}}^{-i\omega t}A(t)dt,
\end{equation}
where $A_{\rm {F}}(\omega)$ represents the Fourier amplitude at frequency $\omega$.

To find excitation of waves, model described in details in \cite{nin13a}, requires applying the FFT on time periods of the same duration before and after the considered perturbation. The first step of this procedure is calculation of ratios of the Fourier amplitude in the time periods "2" $A_F(f;2)$ and "1" $A_F(f;1)$:
\begin{equation}
\alpha_{21}(f)\equiv \frac{A_F(f;2)}{A_F(f;1)}
\label{e01a}
\end{equation}

As in the case of analysis in time domain, the detected variations by this equation can be consequences of different events. For this reason, to extract waves induced by one considered phenomenon, two additional criteria are introduced:

\begin{itemize}
  \item Attenuation of the excited waves in time. This property can be analyzed in the same way as in the procedure for determination of the excited wave frequencies but for the time periods immediately after perturbation (period "2")  and a subsequent period lasting the same amount of time (period "3"). Relevant frequency dependent coefficient is:
      \begin{equation}
        \alpha_{23}(f)\equiv \frac{A_F(f;2)}{A_F(f;3)}
        \label{e01b}
      \end{equation}

  \item Statistic confirmation of wave excitation by considered phenomenon taking more examples into consideration.
\end{itemize}

\section{Results and discussions}
\label{sec:res}

Here, we present detections of SIDs using VLF signal analyses in time and frequency domain. The resulting detections are further connected with our recent research of the solar X-ray flare and GRB events influences on the low ionosphere as well as with an investigation of the possible relationship between earthquakes and the ionospheric  perturbations, for which detailed analyses will be the focus of our upcoming studies.

\subsection{Signal evolution in time domain}

As we said in Sect. \ref{sec:1} the  method for SID detections depends on their characteristics, especially on their intensity, duration, and repetition. Here we give examples for detections of strong and weak SIDs.

\subsubsection{Detection of strong SIDs}

\begin{figure}
\resizebox{0.45\textwidth}{!}{%
  \includegraphics{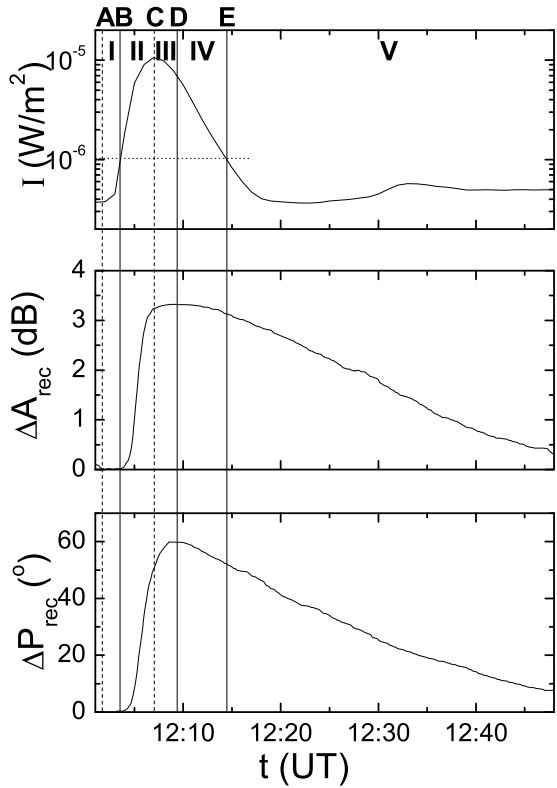}
}
\caption{Time evolutions of the X-radiation flux $I$ in wavelength domain between 0.1 nm and 0.8 nm recorded by the GOES-15 satellite (upper panel), amplitude $\Delta A$ (middle panel) and phase $\Delta P$ (bottom panel) changes of the DHO signal recorded by the AWESOME receiver in Serbia. The vertical lines indicated as A, B, C, D and E indicate start of the time intervals TI I and boundaries between time intervals TI I - TI V described in text. The presented data are collected during influence of the solar X-ray flare occurred on March 24, 2011.}
\label{fig:reznesignali}
\end{figure}

As noted above, when SID is strong enough, we can use a comparison with independent detections of different processes and link it to some event. One of the most important sudden perturbers of the ionospheric D-region plasma is the solar X-ray flare \cite{sch13,sul14,kol14,sul16} and here we will explain the main characteristics of SID on the example of this phenomenon occurred on March 24, 2011. In Fig. \ref{fig:reznesignali} there are presented time evolutions of the X-radiation flux $I$ recorded by the GOES-15 satellite in wavelength range between 0.1 nm and 0.8 nm (upper panel) and  amplitude $\Delta A$ (middle panel) and phase $\Delta P$ (bottom panel) changes of the DHO signal recorded by the AWESOME receiver in Serbia in considered period. For comparison of satellite and VLF receiver recorded variations considering characteristic time periods, we can divide SID period in three specific time domains (TDs):
\begin{itemize}
  \item[$\bullet$] \textbf{TD 1.} In this period denoted as time interval (TI) I the intensity of the X-radiation is too small to induce an increase in the electron production which can noticeably  change receiving radio signal characteristics. For this reason the electron density increase is not detected and we can say that the start of the SID detection has a time delay with respect to the solar X-ray flare detection by satellite.
  \item[$\bullet$]  \textbf{TD 2.} As we presented at the beginning of this section related to the explanation of Fig. \ref{fig:signali3}, the signal variations are different during the period of increased radiation after the SID beginning. Because of that modeling of the D-region plasma is necessary for comparison of variations in the radiation and ionospheric plasma characteristics.
      In the considered case, there are three TIs in this TD: TI II where both the radiation intensity as well as the signal amplitude and phase increase, TI III where $\Delta A$ and $\Delta P$ still grow in spite of start of solar radiation attenuation, and TI IV where signal characteristics also decrease. The last one is the  simplest example where the signal has a similar time evolution as the X-radiation flux with time delay in relevant maximum values.
      However, other events may have different signal variations (see Fig. \ref{fig:signali3}): decrease, complex shape, saturation in the extreme values etc. For these parameters the division of TD 2 is different than in this case.
  \item[$\bullet$]  \textbf{TD 3.} When intensity of the X-radiation decreases to the values which do not produce enough electrons that noticeably affect ionization processes we can assume that the influence of a solar X-ray flare ends. Here we take that this happens when the intensity of the flare falls back to the level where the VLF signal started to increase (borderline between TD 1 and TD 2). Although there are no perturber influences in this TI V, signal characteristics reach values similar to those before the disturbance only at the end of TI V. So we can conclude that there are some relaxation periods to balance the processes of production and loss of electrons.
\end{itemize}

The explained properties related to Fig. \ref{fig:reznesignali} are also noticed in several other events (solar X-ray flares occurred on May 5, 2010, February 18, 2011, April 22, 2011 (two flares)) which we studied in \cite{nin14,baj15,tod16,nin11,nin12b,nin12a,nin15b}.
In these studies we analyzed different space and time dependent plasma parameters in the D-region (the electron density, effective recombination coefficient, electron temperature, electron plasma frequency, contribution of Ly$\alpha$ line in ionization processes, electron content in the D-region and its contribution in changes of the total electron content during solar X-ray flares) using:
\begin{itemize}
  \item The equation for electron density dynamics:
  \begin{equation}
\label{enovo3} \frac{dN(\vec{r},t)}{dt} ={\cal G}(\vec{r},t)-{\cal
L}(\vec{r},t),
\end{equation}
where ${\cal G}(\vec{r},t)$ and ${\cal L}(\vec{r},t)$ are the
electron gain and electron loss rate respectively that are related
to the location $\vec{r}$ and time $t$.
  \item The electron density $N_e$ calculations \cite{tho93} from the Wait's parameters $H^\prime$ and $\beta$ obtained using the LWPC numerical model:
        \begin{equation}
        \label{eq::Ne}
            \displaystyle N_{\rm e}(h,t)
            = 1.43\cdot10^{13}{\rm e}^{-\beta(t)H^\prime(t)}{\rm e}^{(\beta(t)-\beta_0)h},
        \end{equation}
    where $N_e$ is in m$^{-3}$, $H^\prime$(t) and $h$ are in km, $\beta$ is
    in km$^{-1}$ and $\beta_0=0.15$ km$^{-1}$.
  \item Procedure for numerical determination of the best combination of Wait's parameters which satisfied conditions \cite{gru08}:
        \begin{equation}
        \label{eq::uporedjivanjeA}
            \displaystyle \Delta A_{\rm {sim}}(\beta,H')\approx \Delta A_{\rm {rec}}(t),
        \end{equation}
         \begin{equation}
        \label{eq::uporedjivanjeP}
            \displaystyle  \Delta P_{\rm {sim}}(\beta,H')\approx \Delta P_{\rm {rec}}(t).
        \end{equation}
    where $\Delta A_{\rm {sim}}$ and $\Delta P_{\rm {sim}}$ are simulated amplitude and phase changes, while $\Delta A_{\rm {rec}}$ and $\Delta P_{\rm {rec}}$ are registered ones. This procedure explained in \cite{nin14b} shows how we can use VLF/LF recorded data in diagnostic of the D-region plasma.
\end{itemize}


  \begin{figure}
	\resizebox{0.5\textwidth}{!}{%
		\includegraphics{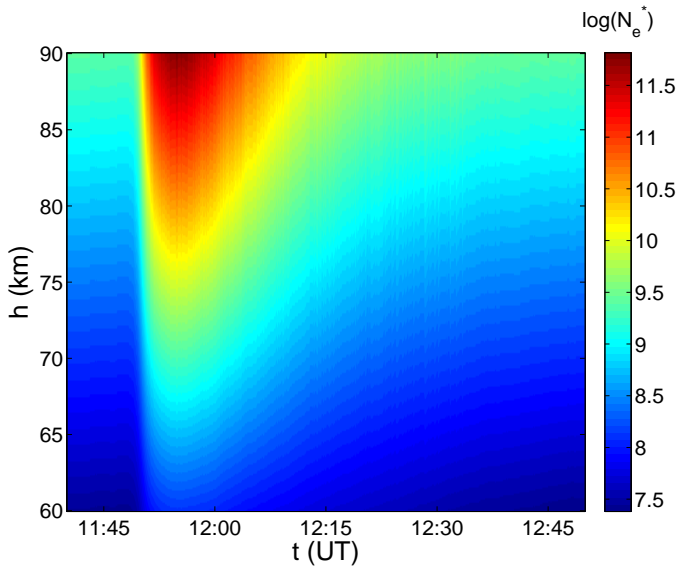}
	}
	\caption{Surface plot of log$(N_e^*(t,h))$ as a function of time $t$ and altitude $h$ during the considered solar X-ray flare where $N_e^*=N_e/N_e^0$, and $N_e^0=1$ m$^{-3}$ \cite{tod16}.}
	\label{fig:Ne}
\end{figure}

 The calculated electron density values (see Fig. \ref{fig:Ne}) show that these examples represent the group of events for which the electron density and signal characteristics have very similar time evolution shapes. For this reason the time intervals and time domains given in Fig. \ref{fig:reznesignali} are relevant for the electron density, too.
 The analytical and numerical procedures given in these studies are based on recorded VLF/LF data and they are developed for calculations of:
 \begin{itemize}
   \item The electron density, electron plasma frequency and index of refractivity during perturbation time period \cite{baj15,nin12a}.
   \item The photo-ionization rate in the upper part of the ionospheric D-region induced by the Ly$\alpha$ line radiation coming from the Sun in unperturbed conditions \cite{nin14}.
   \item The effective recombination coefficient during relaxation period \cite{nin12b}.
   \item The Ly$\alpha$ line contribution in the ionization rate in the maximum of X-radiation flux \cite{nin15b}.
   \item The electron temperature during relaxation period \cite{baj15}.
   \item The D-region electron content contribution in the total electron content \cite{tod16}.
 \end{itemize}

 The main results of presented studies are:
  \begin{itemize}
    \item The existence of a time delay between the onset and the maximum of the electron density perturbation with respect to the corresponding phases of the X-ray flux time evolution and the existence of a relaxation period for the D-region plasma \cite{nin14}.
    \item The dominant influence of the increased intensity of radiation lines in the X-ray spectrum on the enhancement of the electron density in the D-region during the solar flares \cite{nin11,nin15b}.
    \item Increase of the electron density more than one order of magnitude at the top of the D-region \cite{nin14}.
    \item Increase of the effective recombination coefficient at the end of the relaxation and its decrease with altitude \cite{nin12b}.
    \item Increase in contribution of the D-region electron content in the total electron content with X-radiation intensity maximum \cite{tod16}.
    \item Decrease of the electron temperature at the end of the relaxation and its increase with altitude \cite{baj15}.
    \item Increase of the electron temperature changes with altitude at the end of the relaxation \cite{baj15}.
  \end{itemize}

\subsubsection{Detection of weak SIDs}

Detection of a weak SID depends on its characteristics: the intensity, duration and repetition.
In these cases statistical analysis is needed to establish a potential link between some phenomenon and the considered type of SID. Here we present  the methods for examination of the weak VLF/LF signal changes. We analyze periods around GRBs and earthquake.

\begin{description}
  \item[\textbf{Extraction of amplitude peaks.}]
  Fig. \ref{fig:GRBs} presents the increase in number of amplitude peaks $r$ = 2, 3, 4 and 5 times larger than  noise amplitude for a sample of 54 GRBs lasting less than 1 min which are observed by the SWIFT satellite in the period 2009-2012. In this sample we took in consideration the 24 kHz VLF signal emitted by NAA transmitter located in the USA and received in Serbia with time resolution of 20 ms within the periods from 2 min before to 2 min after satellite detection which is not significantly affected by other perturbers. These relatively quiet conditions are required for the shown procedure because of numerous peaks induced by other events which can significantly change statistics, as it is shown in Fig. \ref{fig:nesigurnosti}.

   \begin{figure}
\resizebox{0.5\textwidth}{!}{%
  \includegraphics{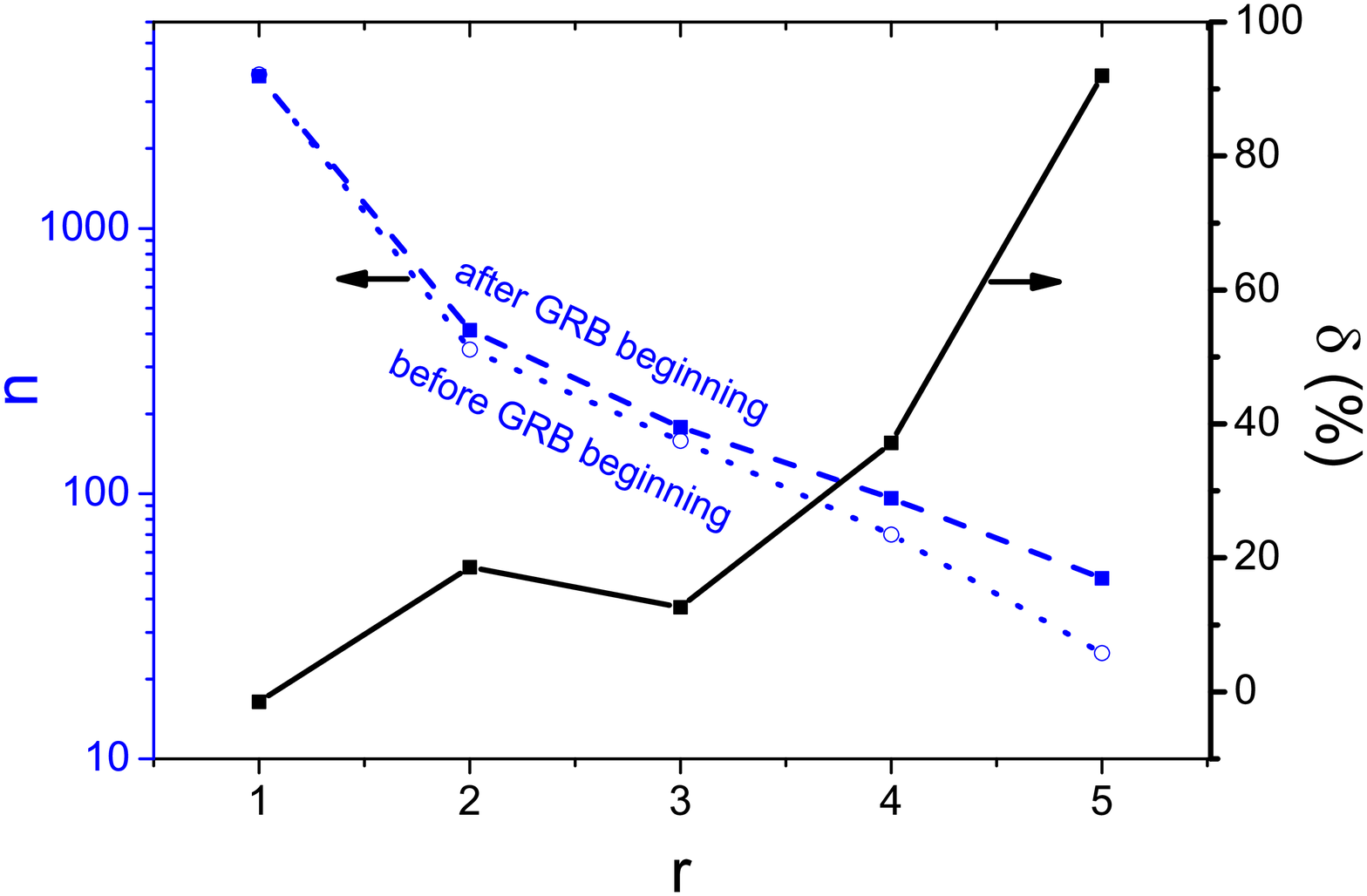}
}
\caption{The number of the relevant amplitude peaks that are $r=1,2,3,4$ and 5 times larger than the noise amplitude of the signal emitted by NAA (the USA) transmitter and received by AWESOME receiver in Serbia in periods of 2 minutes before (dotted line) and after (dashed line)  detections of 54 GRB events by Swift satellite (left y axis) and their increase (solid line) after GRB events detection with respect to period before (right y axis).}
\label{fig:GRBs}
\end{figure}

  As one can see, the increase of number of relevant amplitude peaks is larger in periods of 2 min after GRB detections than in the same duration period before which confirms detectability of a short term reaction of the low ionosphere to GRBs.

   Here we point out that this method is based on the statistical analysis of one type event data set and it gives information about detectability of SIDs that can be connected with the considered phenomenon.
  \item[\textbf{Comparison with the relevant quiet period.}]  As an example for this method we present the comparison of amplitude of the 20.27 kHz VLF signal emitted by the transmitter ICV in Isola di Tavolara, Italy, and received by the AbsPAL receiver located in the Institute of Physics in Belgrade, Serbia for three days in sunset period: the day of the earthquake that struck near Kraljevo, Serbia, on November 3, 2010 (00:57 UT) and one day before and after it. As we can see in Fig. \ref{fig:3dana}, in the day of the earthquake the inverse peaks after 15 h UT, A$_{01}$ and A$_{02}$ start earlier than in the other two days (A$_{-1}$ for day before and A$_{1}$ for day after). It is important to point out that two minima are recorded although the two reference days indicate the existence of just one. Here we can not certainly claim that the signal changes are induced by the earthquake. However it is important to emphasize that similar correlation of the earthquake occurrence with such a signal perturbation is analyzed also in \cite{mol98} and processing of the sample of the signal amplitudes in periods around particular earthquake events can be used for examination of the considered relationship.

\begin{figure}
\resizebox{0.45\textwidth}{!}{%
  \includegraphics{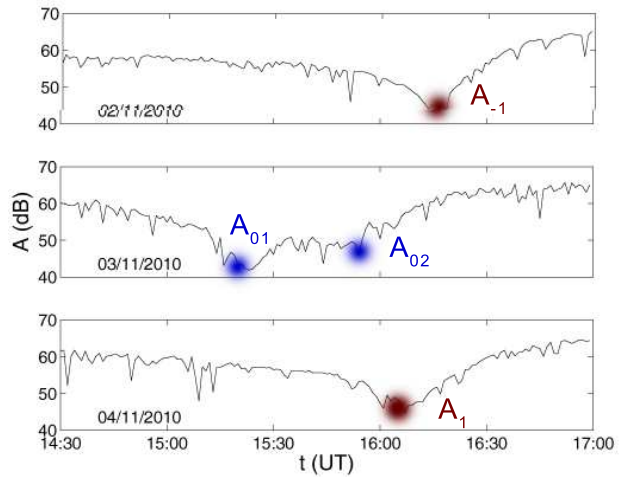}
}
\caption{Amplitudes of the VLF signal emitted in Italy by the ICV transmitter and received in Serbia during the day before (upper panel), the day of the earthquake occurred near Kraljevo, Serbia, on November 3, 2010 (00:57 UT) (middle panel) and the following day (bottom panel). The points indicate the inverse peak minima which are typical of the sunset periods.}
\label{fig:3dana}
\end{figure}
\end{description}

\subsection{Signal evolution in frequency domain}

\begin{figure}
\resizebox{0.45\textwidth}{!}{%
  \includegraphics{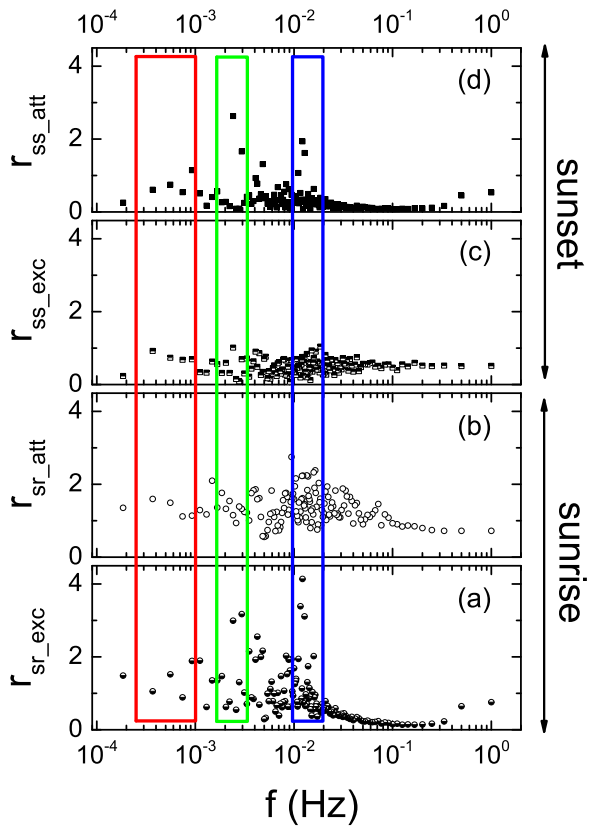}
}
\caption{Coefficients of excitations (panels (a) and (c)) and attenuations (panels (b) and (d)) of the waves for periods of sunrise (bottom panels) and sunset (upper panels) on May 9, 2010. Noticed areas represent frequency domains of the increase in their values e.g. indicate existing of ST induced waves.}
\label{fig:RezAGWkoeficjenti}
\end{figure}

As illustrative example of method based on frequency analysis, we apply the procedure described in Sect. \ref{sec:frequ} on ST induced waves in the low ionosphere. We consider the 90 min time intervals at the beginning and end of the daytime and night sections when quasi-stationary conditions of the basic
state (needed for analysis of the linear waves) are achieved. These intervals are noted as "a" (before sunrise), "b" (after sunrise), "c" (before sunset) and "d" (after sunset). The relevant coefficients related to sunrise $r_{sr_-exc}$ and $r_{sr_-att}$, and sunset $r_{ss_-exc}$ and $r_{ss_-att}$ on May 9, 2010 are calculated by Eqs.  (\ref{e01a}) and (\ref{e01b}), and plotted in Fig. \ref{fig:RezAGWkoeficjenti}.
Here we can indicate three common peak domains in all plots. They
correspond to perturbations that produce much larger Fourier amplitudes after perturbations and much smaller ones at the end than at
the beginning of the related time section. These domains lie between $3\cdot10^{-4}$ Hz and $10^{-3}$ Hz, $3\cdot10^{-4}$ Hz and $4\cdot10^{-3}$ Hz and $10^{-2}$ Hz and $2\cdot10^{-2}$ Hz. According to the analysis given in Sect. \ref{AGWtheory} the first two oscillation frequency domains correspond to the gravity modes and the third one to the acoustic modes. The detailed analysis of five days given in \cite{nin13a} confirms that corresponding wave time periods are a result of the ST induced perturbation which is in agreement with those for higher altitudes of the E- and F-regions
from literature \cite{her06,afr08}. Also, similar fluctuations in the
form of magnetohydrodynamic waves were also found in
high magnetospheric regions as externally driven modes
with typical periods ranging from few seconds to more than
1000 s \cite{dek01a,dek01b}.

\section{Summary}
\label{sec:sum}

To conclude this research we point out the existence of different SIDs which required different methods for their detections by VLF/LF radio signals used for the low ionospheric monitoring. Here we give the classification of the SIDs and suggest several methods for their detections using time domain analyses:

\begin{itemize}
  \item Strong SIDs. These plasma perturbations are sufficiently large to cause detectable changes in signal characteristics and in these cases we can analyze relationship between one particular event properties, and the ionosphere and, consequently, VLF/LF signals reactions. Here we point out time delays in the ionospheric response and its relaxation period after perturbation. Also, we point out and explain differences in the signal reactions to some phenomena.
  \item Weak SIDs. The uncertainty in detections of these particular events is induced by:
      \begin{itemize}
        \item Weak intensity of SIDs;
        \item The occurrence of other nearly simultaneous phenomena that induce very similar variations to the ones sought for;
        \item No clearly visible changes in recorded signal properties.
      \end{itemize}
      Here, statistical analysis is needed to establish a potential link between a phenomenon and the considered type of SID. The suggested methods which can be used for possible confirmation of the low ionospheric reactions to the considered phenomenon are:
      \begin{itemize}
        \item Extraction of amplitude peaks.
        \item Comparison with the relevant quiet period.
        \item Superposed epoch technique.
      \end{itemize}
\end{itemize}

In addition to the explained procedure for signal analyses in the time domain, we present a method for detection of AGWs. It is based on implementation of the FFT on recorded signal amplitude within periods before, immediately and some time after several  events of the same type in order to determine the excited wave frequencies, their attenuation and repetition, which, consequently, confirm induction of AGWs by the considered phenomenon.

All these procedures have been described with some relevant examples for better clarity.

In addition to the application of various experimental settings to monitor different height domains, there are several types of measurements of the same area. Each of them has its own advantages and disadvantages. The presented study shows that our experimental equipment is completely suitable for the monitoring of different SIDs: periodical and unperiodical, long-lasting and short-lasting, global and local, strong and weak. This is possible since there is a continuous emission and reception of radio signals, very good time resolutions of collected data and numerous worldwide located transmitters and receivers. The main disadvantage of this technique is the absence of information on local plasma medium which can be obtained using in site measurements by the rockets. However, that kind of measurement is not continuous and usually it can not be used for detection of unpredicted SIDs like those induced by e.g. GRBs.

Finally, we want to point out the importance of large databases in different statistical analyses and their combination in statistical analyses of the relatively rare events, like earthquakes, and comparative analyses of the different signals or geographical areas reactions.

%

\section*{Acknowledgements}
This work is supported by the Ministry of Education, Science and Technological Development of the Republic of Serbia under the grants III 44002, 176001, 176002 and 176004. Also, this study is made within the COST projects ES1401 and TD1403 and VarSITY project. We are grateful to the anonymous referees for comments and suggestions that significantly improved our paper. The data for this paper collected by GOES-15 satellite
is available at NOAA's National Centers for Environmental information (\url{http://satdat.ngdc.noaa.gov/sem/goes/data/new_full/2011/03/goes15/csv/g15_xrs_2s_20110324_20110324.csv}). Requests for the VLF data used for analysis can be directed to the corresponding author.

%
 \bibliographystyle{epj}

%

\end{document}